\documentstyle[aps,prb,graphicx]{revtex}
\begin{document}
\title{Self-consistent calculation of the
autolocalization barrier for quasiparticles in
anisotropic crystal.}
\author
{A.Shelkan}
\address
{Institute of Physics, University of Tartu,
Riia 142, EE2400 Tartu, Estonia}

\date{\today}

\maketitle

\begin{abstract}
\small  
The energy of the electron wave packet
interacting with lattice distortion, is considered 
in anisotropic crystal.
Anisotropy of the electron and phonon spectra
as well as of the electron-phonon interaction
are taken into account.
The height of the barrier between
free and self-trapped states is calculated in
dependence on the anisotropy parameters.
The calculation is done numerically, using continual 
aproximation. The analytical solution is obtained for some
cases of quasi-two and quasi-one-dimensional spectra. \\
Key words: anisotropy; polarons; barrier.

\end{abstract}

\pacs{PACS numbers: }



There is a general agreement that in systems with
distortion-type electron-phonon coupling,
an electron (hole, exciton) forms
the self-trapped (polaronic-type) state.
Usually, this state is separated from the free one by
an autolocalization barrier (AB , see \cite{luwik,rashba}).
The theory of the AB has been  developed for isotropic systems
\cite{rashba}. During the last years polarons in
anisotropic systems \cite{belenki}, especially in
superconducting materials \cite{polarons},
are widely discussed. Therefore it is of interest 
to consider how the lattice anisotropy  
influences the AB . Briefly this problem was considered
in \cite{rashba}, where the one- and two-
dimensional limits were studied.
The phonon spectrum and electron-phonon
coupling have been taken to be isotropic.
It was shown, that within these limits
the height of the barrier
$E \sim (m_1m_2m_3)^{-1}$, 
i.e barrier vanishes in one- and two-dimensional cases.
However, in three-dimensional
crystals with finite anisotropy 
the barrier has not been studied.

In this paper, we apply the adiabatic
theory of AB, developed in
\cite{zavt}, to anisotropic crystals.
To make the problem easier
we consider the crystal with symmetry not
lower than romboedrical (i.e. with diagonal tensors
of effective mass and elastic constant). 
To simplify the operator of the electron-phonon
interaction we suppose the localization to be connected 
with the displacement of the particles along
only one direction (signed $z$), that corresponds
to the direction of the lowest dispersion of electron zone.
This approximation is justified by the fact 
that in the wide zone limit considered here,
the influence of the electron-phonon 
interaction on the energy of the electron is inverse
proportional to the width of the zone.
Besides, we suppose that the 
energy of the AB 
essentially exceeds the average energy of acoustical phonons
and the radius of the barrier state is much larger than the
lattice constant. These assumptions allow us to use the 
adiabatic approximation and to solve the problem in the 
continual limit.

The energy functional reads  \cite{zavt}: \ 
$ {\cal E} = {\cal E}_e + {\cal E}_p + V  $

where

\[ {\cal E}_e = - \int d^3 r C(r) \lbrack
J_1 \frac {\delta ^2} {\delta x^2} +
J_2 \frac {\delta ^2} {\delta y^2} +
J_3 \frac {\delta ^2} {\delta z^2} \rbrack C(r), \]

\begin{equation}
{\cal E}_p = \int d^3 r \lbrack
\beta _1 \left( \frac {\delta u (r)} { \delta x} \right) ^2 +
\beta _2 \left( \frac {\delta u (r)} { \delta y} \right) ^2 +
\beta _3 \left( \frac {\delta u (r)} { \delta z} \right) ^2 \rbrack ,
\end{equation}

\[ V = - \int d^3 r \lbrack \lambda
\frac {\delta u (r)} { \delta z} C^2(r) \rbrack . \]
Here $C(r)$ denotes the wave function, and
${\cal E}_e$ - the cinetic energy of the electron,
${\cal E}_p$ is the lattice elastic energy,
$ V $ - the energy of the electron-phonon coupling , and
$u(r)$ labels the component of the atomic displacement.
In this model exists a self-consistent state
that corresponds to the barrier and
that is the saddle point of the energy functional .
To determine this point we vary the energy functional
initially with respect to $u(r)$.
Then we choose the wave function $C(r)$ in
exponential form
\begin{equation}
C(r) = \left( \frac {8} {\pi ^3} \right) ^{1/4}
(R_1R_2R_3)^{-1/2} \exp \{
-\frac {x^2} {R_1^2}
-\frac {y^2} {R_2^2}
-\frac {z^2} {R_3^2} \} ,
\end{equation}
and consider $R(i)$ as variational parameters,
corresponding to the radius of the electron state.
As a result we get for the energy functional
\begin{equation}
{\cal E} = \sum _{i=1}^3 \frac {J_i} {R_i^2} -
\frac {\lambda^2} {4} \pi f(R_1,R_2,R_3),
\end{equation}
where
\begin{equation}
f(R_1,R_2,R_3) = \int_0^{\infty } dt \, t^2 \lbrack
(\beta _1 + R_1^2 t^2)
(\beta _2 + R_2^2 t^2)
(\beta _3 + R_3^2 t^2)^3
\rbrack ^{-1/2} .
\end{equation}

For reducing the number of free parameters we simplify
the problem: we restrict ourselves
to a system that is isotropic in the
$(x,y)$ plane, i.e. we suppose that
$ R_1 = R_2 =R ; \ J_1 = J_2 = J ; \ \beta _1 = \beta _2 = \beta . $
Parameters $\beta $ correspond to the elastic constants
\cite{landau}:
$\beta = \lambda_{zxzx}$ ;
$\beta _3 = \lambda_{zzzz}$ .
In the isotropic systems
$\lambda_{zzzz} -
\lambda_{zzxx} -
2 \lambda_{zxzx} = 0$.
In the  approximation of central forces
$\lambda_{zxzx} = \lambda_{zzxx}$, 
and the elastic isotropy condition takes the form
$\beta _3 = 3 \beta $.
Evidently the condition of spatial isotropy 
for the electron spectrum is $J_3 = 3 J $ .
It is convenient to characterize the violation of 
spatial isotropy by the dimensionless parameters
$ \gamma = J / J_3 \ , \ 
\alpha = 3 \beta / \beta _3 . $

Then
\begin{equation}
{\cal E } = J_3 \lbrack
\frac {2 \gamma} {R^2} +
\frac {1} {R_3^2} -
\frac {\lambda ^2 \pi } {4 J_3 \beta _3 }
\int dx \frac {x^2}
{ \left( \frac {\alpha } {3} \frac {R_3^2} {R^2} + x^2 \right)
(1 + x^2)^{3/2} } \rbrack .
\end{equation}
Let us turn to the new variational parameters
$d$ and $\theta $, introduced according to:
$ R^2 = d\sin \theta , R_3^2 = d \cos \theta$.
After variation with repect to  $d$ the energy functional 
takes the form:
\begin{equation}
{\cal E } =
\frac {64} {27} J^3 \beta ^2 \lambda ^{-4} ctg \, \theta
\frac {(2 \gamma + tg \, \theta )^3 }
{\lbrack \int dx
\frac {x^2}
{(\alpha \, ctg \theta + x^2 ) ( 1 + x^2 )^{3/2} } \rbrack ^2 } .
\end{equation}
Now the problem is reduced to the minimization
of the energy with respect to one single parameter $\theta $ - 
this minimal value $ E $ is the height of the barrier.
We perform this minimization numerically.

From (6) one can see that, besides $\alpha $ and $\gamma $ ,
the result depends also on $J , \beta $ and $\lambda $. 
Our goal is to elucidate 
how the anisotropy parameters
influence the height of the barrier.
For this purpose we shall look for the ratio $E/E_0 $ of the 
heights in anisotropic and isotropic cases
$E/E_0$. 
For the  scale parameters it is convenient to choose
$ J^{is} = J_3^{is} = (2 J + J_3) / 3 , $
$ \beta ^{is} = \beta _3^{is} = (2 \beta  + \beta _3) / 3 , $
$ \lambda _{is} = \lambda . $
Then the ratio $E/E_0$ depends
only on the anisotropy parameters $\alpha $ and $\gamma $ .
Similarly we introduce the ratios $R/R_0$ and $R_3/R_0$.
The results of numerical calculations of the ratios
$E/E_0 $
as functions of the anisotropy parameters
are presented in Figs 1-2.
One can see that the maximal height of AB 
corresponds to $\gamma \sim 1$ and $\alpha \sim 0.1$.
In the case of the quasi-one-dimensional ($\gamma << 1$) 
electron spectrum barrier vanishes slower than in
the quasi-two-dimensional limit ($\gamma >> 1$) .
The transition to the one-dimensional phonon spectrum
($\alpha << 1$)
only weakly affects the height of the barrier,
and for large  $\gamma $ doesn't affect the barrier at all.


We note that for the parameters of anisotropy,
being used in our numerical calculations, the ratios
$R/R_0$ and $R_3/R_0$ are of the order of one or larger.
Therefore the continual approximation yields a correct description
of the AB in anisotropic crystals if it is applicable
in corresponding isotropic crystals.
For $\alpha >> \gamma $ and $\gamma \rightarrow 0 $
the ratio $ R/R_0 \rightarrow 0 $ , i.e. in this limit we
cannot use the continual approximation.
Besides, in anisotropic cases, i.e. when AB vanishes,
then either one or both radii go to infinity.

In the limiting cases of large and small values of
the ratio $\alpha / \gamma $
one can obtain analytic solutions.
Denoting
$\alpha R_3^2 / 3 R^2$ as $u$,
the integral in (5) as 
$f(u)$ and using
formula 3.197.1 from
\cite{gradshtein} 
one obtains
\begin{equation}
f(u) = \int dx \frac {x^2} {(u + x^2) (1 + x^2)^{3/2}} =
1 - \frac 2 3 F \left( 1, \frac 1 2 ; \frac 5 2 ; 1 - u^{-1} \right) ,
\end{equation}
where $F$ is the hypergeometric function.

Let us consider first the case
$\alpha << \gamma $ and assume that $u << 1$. 
Transforming (7) according to 9.132.1 from \cite{gradshtein},
neglecting the second order terms in  $u^{1/2}$ 
and variating with respect to  $R$ and $R_3$, we obtain
\[ \frac {E} {E_0} \sim
\frac {\gamma ^2} { ( 6 \alpha + 1)^2 (2 \gamma + 1 )^3}
\left( 1 + 4 \sqrt {\frac {\alpha} {2 \gamma}} \right) ,\]
\begin{equation}
\frac {R} {R_0} \sim
\frac { ( 6 \alpha + 1) (2 \gamma + 1 )} {\gamma ^{1/2}}
\left( 1 - \frac 3 2 \sqrt {\frac {\alpha} {2 \gamma}} \right) ,
\end{equation}

\[ \frac {R_3} {R_0} \sim
\frac { ( 6 \alpha + 1) (2 \gamma + 1 )} {\gamma }
\left( 1 - 3 \sqrt {\frac {\alpha} {2 \gamma}} \right) .\]
From (8) one can see, that $u \sim \alpha / \gamma $ . 
The condition $u << 1$ is fulfilled, so
the corresponding assumption is justified.

Now let us consider the case $\alpha >> \gamma $.
Here we assume that $u >> 1$. 
Using the properties of hypergeometric function 
we rewrite (7) as
\begin{equation}
f(u) = \frac 1 3
F \left( 1, 1; \frac 5 2 ; 1 - u \right)
\end{equation}
Neglecting the second order terms in $1/u$ and
variating with respect to $R$ and $R_3$ we get
\[ \frac {E} {E_0} \sim
\frac {\alpha ^2} { ( 6 \alpha + 1)^2 (2 \gamma + 1 )^3}
\left( \ln {\frac {\alpha} {\gamma}} \right) ^{-2} ,\]
\begin{equation}
\frac {R} {R_0} \sim
(2 \gamma + 1) \gamma ^{1/2}
\left( \ln \frac {\alpha} {\gamma} \right) ^{-3/2} ,
\end{equation}

\[ \frac {R_3} {R_0} \sim
(2 \gamma + 1) \ln \frac {\alpha} {\gamma} . \]
One can see from (10) , that the condition $u >> 1$ is fulfilled.
In the case of the quasi-two-dimensional phonon spectrum $(\alpha >> 1)$,
$E/E_0$ goes to zero as $(\ln \alpha )^{-2}$,
and  $R/R_0$ and $R_3 / R_0$ goes to infinity 
correspondingly as
$(\ln \alpha )^{3/2}$ and $\ln \alpha $ . 
The transition to the quasi-one-dimensional phonon spectrum
$(\alpha << 1)$ 
slightly influences the height of the AB.
If the electron spectrum is quasi-two-dimensional
$(\gamma >> 1)$, then
$E/E_0$ goes to zero as $1 / \gamma $,
and $R/R_0$ goes to infinity as $\gamma ^{1/2}$ .
This means, that the effect of anisotropy of
the electronic  spectrum to the AB is much stronger,
than of the anisotropy of the phonon spectrum.

Note that within the framework of this model, 
the consideration of the limit, which correponds 
to the one-dimensional electron spectrum $(\gamma << 1)$
is not justified. 
Indeed, when $\gamma $ goes to zero,
then as one can see from (12), the ratio $R/R_0$ also goes
to zero and the continual approximation becomes
inapplicable.

In conclusion, in this communication the influence of the
anisotropy of electron and phonon spectra
and also of electron-phonon interaction on the height
of the barrier of a phonon polaron is investigated.
It is shown, that the anisotropy of
the electron spectrum stronger affects the AB,
than the anisotropy of the phonon spectrum.
In the quasi-two-dimensional limit
the height of the barrier goes to zero.
We emphasize that the disapearence of the barrier
in the quasi-two-dimensional case is characteristic
for the phonon polaron.
On the
contrary, as it was pointed out in \cite{newst}, a spin-polaron model
characterized is by the a finite barrier also in a two-dimensional 
AF lattice.
The reason of this difference between
phonon and spin polarons
lies in the strongly nonlinear nature of the localized
spin excitation: one turned spin forms a stable topological
defect in the AF lattice even without a charge carrier.
Therefore a large-size wave-packet
of a free particle with one turned spin represents a
metastable state ( unlike the wave packet with
a local lattice distortion, which is fully unstable already
in the time scale of the order of the vibrational period ).
This difference in the properties of the phonon and spin 
polarons must be taken into account
when discussing the transport properties of the 
high-$T_c $ superconductors.

I wish to express my gratitude to D.Nevedrov
for valuable discussions and stimulating of
this publication.

\newpage

\begin{figure}
\begin{center}
\includegraphics{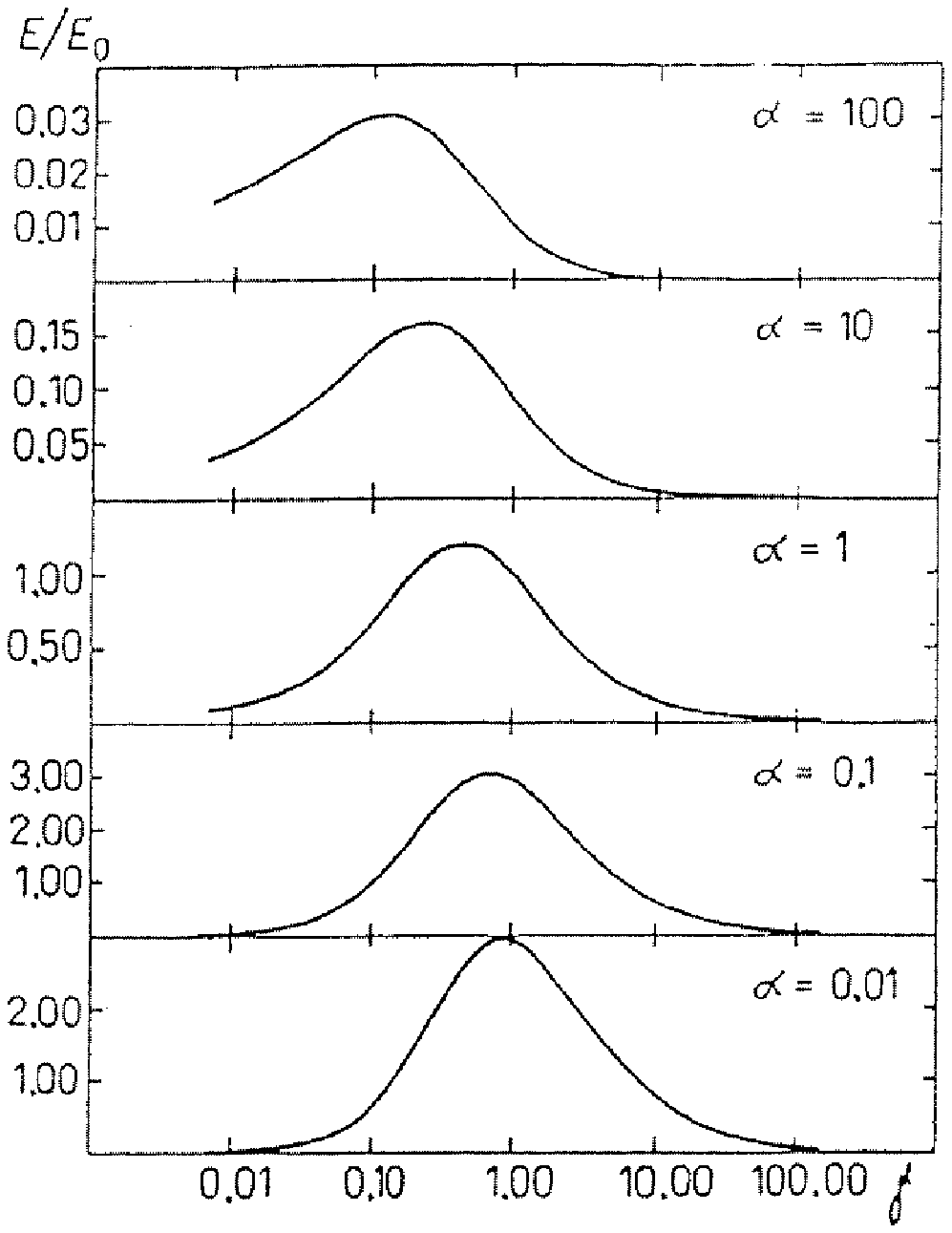}
\newline 
{\bf Fig.1} The ratio of the heights of the barrier in
anisotropic and isotropic cases $E/E_0$
vs anisotropy parameter of electron spectrum $\gamma $
for various anisotropy parameters of phonon spectrum 
$\alpha $.
\end{center}
\end{figure}

\begin{figure}
\begin{center}
\includegraphics{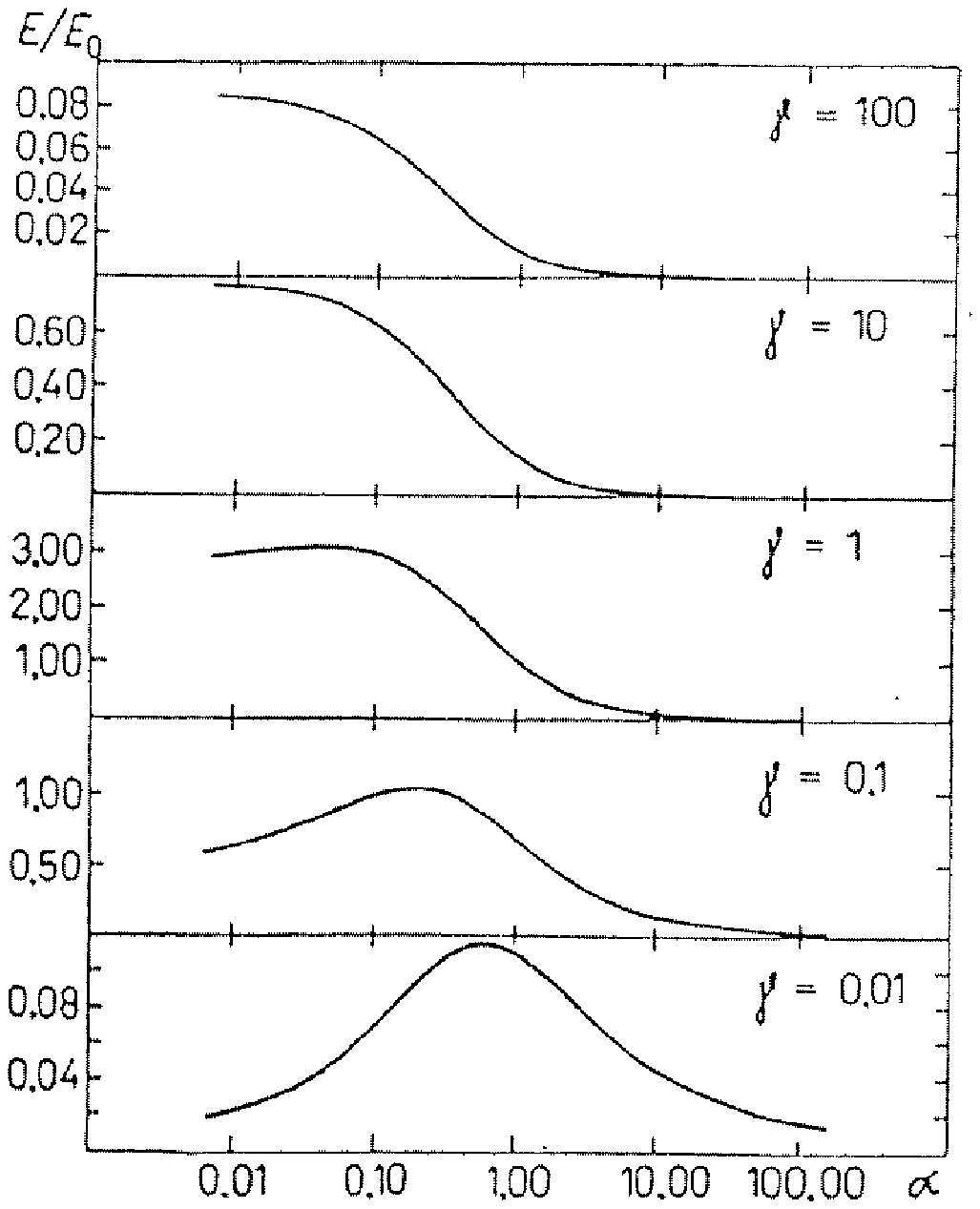}
\newline 
{\bf Fig.2} The ratio of the heights of the barrier in
anisotropic and isotropic cases $E/E_0$
vs anisotropy parameter of phonon spectrum $\alpha $
for various anisotropy parameters of electron spectrum 
$\gamma $.
\end{center}
\end{figure}

\end{document}